\documentclass[preprint,showpacs,preprintnumbers,amsmath,amssymb]{revtex4}
\usepackage{graphicx}% Include figure files
\usepackage{dcolumn}% Align table columns on decimal point
\usepackage{bm}% bold math

\newcommand{\be}{\begin{equation}}
\newcommand{\en}{\end{equation}}
\newcommand{\bea}{\begin{eqnarray}}
\newcommand{\ena}{\end{eqnarray}}

\begin{document}

%\preprint{GACG/07/2006}

\title{ Warm-Intermediate  inflationary universe model  }

\author{Sergio del Campo}
 \email{sdelcamp@ucv.cl}
\affiliation{ Instituto de F\'{\i}sica, Pontificia Universidad
Cat\'{o}lica de Valpara\'{\i}so, Casilla 4059, Valpara\'{\i}so,
Chile.}
\author{Ram\'on Herrera}
\email{ramon.herrera@ucv.cl} \affiliation{ Instituto de
F\'{\i}sica, Pontificia Universidad Cat\'{o}lica de
Valpara\'{\i}so, Casilla 4059, Valpara\'{\i}so, Chile.}

\date{\today}% It is always \today, today,

\begin{abstract}
  Warm inflationary universe models in the context of  intermediate
 expansion, between power law and exponential, are studied.
 General  conditions required for these models
 to be realizable are derived and
 discussed. This study is done in the weak and strong dissipative
 regimes. The inflaton potentials considered in this study are negative-power-law and
 powers of logarithms, respectively. The parameters of our models are constrained from the
 WMAP three and five year data.
\end{abstract}

\maketitle

\section{Introduction}

It is well  know that warm inflation, as opposed to the
conventional cool inflation, presents the attractive feature that
it avoids the reheating period \cite{warm}. In these kind of
models dissipative effects are important during the inflationary
period, so that radiation production occurs concurrently together
with the inflationary expansion. If the radiation field is in a
highly excited state during inflation, and this has a strong
damping effect on the inflaton dynamics, then, it is found a
strong regimen of  warm inflation. Also, the dissipating effect
arises from a friction term which describes the processes of the
scalar field dissipating into a thermal bath via its interaction
with other fields. Warm inflation shows how thermal fluctuations
during inflation may play a dominant role in producing the initial
fluctuations  necessary for Large-Scale Structure (LSS) formation.
In this way, density fluctuations arise from thermal rather than
quantum fluctuations \cite{62526}. These fluctuations have their
origin in the hot radiation and influence the inflaton through a
friction term in the equation of motion of the inflaton scalar
field \cite{1126}. Among the most attractive features of these
models, warm inflation end  at the epoch when the universe stops
inflating and "smoothly" enters in a radiation dominated Big-Bang
phase\cite{warm}. The matter components of the universe are
created by the decay of either the remaining inflationary field or
the dominant radiation field \cite{taylorberera}.

A possible evolution during inflation is the particular scenario
of intermediate inflation, in which the scale factor, $a(t)$,
evolves as $a=\exp(A t^f)$, where $A$ and $f$ are two constants,
where $0<f<1$; the expansion of this universe is slower than
standard de Sitter inflation ($a=\exp(H t)$), but faster than
power law inflation ($a= t^p; p>1$), this is the reason why it is
called "intermediate". This model was introduced as an exact
solution for a particular scalar field potential of the type
$V(\phi)\propto \phi^{-4(f^{-1}-1)}$\cite{Barrow1}. In the
slow-roll approximation, and with this sort of potential, it is
possible to have a spectrum of density perturbations which
presents a scale-invariant spectral index, i.e. $n_s=1$, the
so-called Harrizon-Zel'dovich spectrum provided that $f$ takes the
value of $2/3$\cite{Barrow2}. Even though this kind of spectrum is
disfavored by the current WMAP data\cite{astro,astro2}, the
inclusion of tensor perturbations, which could be present at some
point by inflation and parametrized by the tensor-to-scalar ratio
$r$, the conclusion that $n_s \geq 1$ is allowed providing  that
the value of $r$ is significantly nonzero\cite{ratio r}. In fact,
in Ref. \cite{Barrow3} was shown that the combination $n_s=1$ and
$r>0$ is given by a version of the intermediate inflation  in
which the scale factor varies as $a(t)\propto e^{t^{2/3}}$ within
the slow-roll approximation.

The main motivation to study this sort of model becomes from
string/M-theory. This theory suggests that in order to have a
ghost-free action high order curvature invariant corrections to
the Einstein-Hilbert action must be proportional to the
Gauss-Bonnet (GB) term{\cite{BD}}. GB terms arise naturally as the
leading order of the $\alpha$ expansion to the low-energy string
effective action, where $\alpha$ is the inverse string
tension{\cite{KM}}. This kind of theory has been applied to
possible resolution of the initial singularity
problem{\cite{ART}}, to the study of Black-Hole solutions{\cite{
Varios1}}, accelerated cosmological solutions{\cite{ Varios2}}. In
particular , very recently, it has been found{\cite{Sanyal}} that
for a dark energy model the GB interaction in four dimensions with
a dynamical dilatonic scalar field coupling leads to a solution of
the form $a = a_0 \exp{A t^{f}}$,  where the universe starts
evolving with a decelerated exponential expansion. Here, the
constant $A$ becomes given by $A= \frac{2}{\kappa n}$ and
$f=\frac{1}{2}$, with $\kappa^2 = 8 \pi G$ and $n$ is a constant.
In this way, the idea that inflation, or specifically,
intermediate inflation, comes from an effective theory at low
dimension of a more fundamental string theory is in itself very
appealing.

Thus, our aim in this paper is to study an evolving intermediate
scale factor in the warm inflationary universe scenario. We will
do this for two regimes; the weak and the strong dissipative
regimes.

The outline of the paper is a follows. The next section presents a
short review of  the modified  Friedmann equation and the
warm-intermediate inflationary phase.  In the Sections
\ref{section2} and \ref{section3} we discuss the weak and strong
dissipative regimens, respectively. Here, we give explicit
expressions for the dissipative coefficient, the scalar power
spectrum and the tensor-scalar ratio. Finally, our conclusions are
presented in Section\ref{conclu}.  We chose units so that
$c=\hbar=1$.

\section{The  Warm-Intermediate Inflationary phase.\label{secti} }

We start by writing down  the modified Friedmann equation, by
using the FRW metric. In particular, we assume that the
gravitational dynamics  give rise to a  Friedmann equation of the
form
\begin{equation}
H^2=\frac{\kappa}{3}\,[\rho_{\phi}+\rho_\gamma], \label{HC}
\end{equation}
where $\kappa=8\pi G=8\pi/m_p^2$ (here $m_p$ represents the Planck
mass), $\rho_\phi=\dot{\phi}^2/2+V(\phi)$,
  $V(\phi)=V$ is
the scalar   potential and  $\rho_\gamma$ represents  the
radiation energy density.

 The dynamics of the
cosmological model, for $\rho_\phi$ and $\rho_\gamma$ in the warm
inflationary scenario is described by the equations
 \be\dot{\rho_\phi}+3\,H\,(\rho_\phi+P_\phi)=-\Gamma\;\;\dot{\phi}^2, \label{key_01}
 \en
and \be \dot{\rho}_\gamma+4H\rho_\gamma=\Gamma\dot{\phi}^2
.\label{3}\en Here $\Gamma$ is the dissipation coefficient and it
is responsible of the decay of the scalar field into radiation
during the inflationary era. $\Gamma$ can be assumed to be a
constant or a function of the scalar field $\phi$, or the
temperature $T$, or both \cite{warm}.  On the other hand, $\Gamma$
must satisfy  $\Gamma>0$ by the Second Law of Thermodynamics. Dots
mean derivatives with respect to time.

During the inflationary epoch the energy density associated to the
scalar field dominates over the energy density associated to the
radiation field\cite{warm,62526}, i.e. $\rho_\phi>\rho_\gamma$,
the Friedmann equation (\ref{HC})  reduces  to
\begin{eqnarray}
H^2\approx\frac{\kappa}{3}\,\rho_\phi,\label{inf2}
\end{eqnarray}
and  from Eqs. (\ref{key_01}) and (\ref{inf2}), we can write
\begin{equation}
 \dot{\phi}^2= -\frac{2\,\dot{H}}{\kappa\,(1+R)},\label{inf3}
\end{equation}
where $R$ is the rate defined as
\begin{equation}
 R=\frac{\Gamma}{3H }.\label{rG}
\end{equation}
For the  weak (strong) dissipation  regime, we have $R< 1$ ($R>
1$).

We also consider that  during  warm inflation the radiation
production is quasi-stable\cite{warm,62526}, i.e.
$\dot{\rho}_\gamma\ll 4 H\rho_\gamma$ and $
\dot{\rho}_\gamma\ll\Gamma\dot{\phi}^2$.  From Eq.(\ref{3}) we
obtained that the energy density of the radiation field becomes
 \begin{equation}
\rho_\gamma=\frac{\Gamma\dot{\phi}^2}{4H}=-\frac{\Gamma\,\dot{H}}{2\,\kappa\,H\,(1+R)},\label{rh}
\end{equation}
which  could be written as $\rho_\gamma= C_\gamma\, T^4$, where
$C_\gamma=\pi^2\,g_*/30$ and $g_*$ is the number of relativistic
degrees of freedom. Here $T$ is the temperature of the thermal
bath.

From Eqs.(\ref{inf3}) and (\ref{rh}) we get that
\begin{equation}
T= \left[-\frac{\Gamma\,\dot{H}}{2\,\kappa\,\,C_\gamma
H\,(1+R)}\right]^{1/4}.\label{rh-1}
\end{equation}

From first principles in quantum field theory the dissipation
coefficient $\Gamma$ is computed for models in cases of
low-temperature regimes\cite{26} (see also Ref.\cite{28}). Here,
was developed the dissipation coefficients in supersymmetric
models which have an inflaton together with multiplets of heavy
and light fields. In this approach, it was used   an interacting
supersymmetric theory, which has three superfields $\Phi$, $X$ and
$Y$ with a superpotential, $W=g\Phi X^2+hXY^2$, where $\phi$,
$\chi$ and $y$ refer to their bosonic component. The inflaton
field couples to heavy bosonic field $\chi$ and fermions
$\psi_\chi$, obtain their masses through couplings to $\phi$,
where $m_{\psi_\chi}=m_\chi=g\phi$. In the low -temperature
regime, i.e. $m_\chi,m_{\psi_\chi}>T>H$, the dissipation
coefficient, when $X$ and $Y$ are singlets, becomes \cite{26}
\begin{equation}
\Gamma\simeq0.64\,g^2\,h^4\left(\frac{g\,\phi}{m_\chi}\right)^4\,
\frac{T^3}{m_\chi^2}.\label{G0}
\end{equation}
This latter equation can be rewritten as
\begin{equation}
\Gamma\simeq C_\phi\,\frac{T^3}{\phi^2},\label{G}
\end{equation}
where $C_\phi=0.64\,h^4\,\cal{N}$. Here ${\cal{N}}={\cal{N}}_\chi
{\cal{N}}_{decay}^2$, where $\cal{N}_\chi$ is the multiplicity of
the $X$ superfield and ${\cal{N}}_{decay}$ is the number of decay
channels available in $X$'s decay\cite{26,27}.

From Eq.(\ref{rh-1}) the above equation becomes
\begin{equation}
\Gamma^{1/4}\,(1+R)^{3/4}\simeq\left[\frac{-2\,\dot{H}}{9\,\kappa\,
C_\gamma\,H}\right]^{3/4}\,\frac{C_\phi}{\phi^2},\label{G1}
\end{equation}
which determines the dissipation coefficient in the weak (or
strong) dissipative regime  in terms of scalar field $\phi$ and
the parameters of the model.

 In general the scalar potential can be obtained from
Eqs.(\ref{HC}) and (\ref{rh}) 
\begin{equation}
V(\phi)=\frac{1}{\kappa}\left[3\,H^2+\frac{\dot{H}}{(1+R)}\,\left(1+\frac{3}{2}\,R\right)\right],\label{pot}
\end{equation}
which could be expressed explicitly in terms of the scalar field,
$\phi$, by using Eqs.(\ref{inf3}) and (\ref{G1}), in the
 weak (or strong) dissipative regime.

Solutions can  be found for warm-intermediate inflationary
universe models where  the scale factor, $a(t)$, expands as
follows
\begin{equation}
a(t)=\exp(\,A\,t^{f}).\label{at}
\end{equation}
Recalled,  that $f$ is a dimensionless constant parameter with
range $0<f<1$, and $A>0$ has dimension of $m_p^f$. In the
following, we develop models for a variable dissipation
coefficient $\Gamma$, and  we will restrict ourselves to the weak
(or strong ) dissipation regime, i.e. $R<1$ (or $R>1$).

\section{ The  weak dissipative regime.\label{section2}}

Assuming that, once the system evolves according to the weak
dissipative regime, i.e. $\Gamma<3H$, it remains in such limit for
the rest of the evolution. From Eqs.(\ref{inf3}) and (\ref{at}),
we obtained a relation between the scalar field and cosmological
times given by
\begin{equation}
\phi(t)=\phi_0+\sqrt{\frac{8\,A\,(1-f)}{f\,\kappa}}\,\;t^{f/2},\label{wr1}
\end{equation}
where $\phi(t=0)=\phi_0$.  The Hubble parameter as a function of
the inflaton field, $\phi$, results in
\begin{equation}
H(\phi)=A^{1/f}\,f^{(2f-1)/f}\,\left[\frac{\kappa}{8\,(1-f)}\right]^{(f-1/f)}\,(\phi-\phi_0)^{2(f-1)/f}.\label{HH}
\end{equation}
Without loss of generality $\phi_0$ can be taken to be zero.

From Eq.(\ref{G1}) we obtain for the dissipation coefficient as
function of scalar field
\begin{equation}
\Gamma(\phi)=B\,\phi^{-\beta_1},
\end{equation}
where
$$
B=\frac{8\,C_\phi^4\,(1-f)^3}{27^2\,\kappa^3\,C_\gamma^3}\;\left[\frac{8\,A\,(1-f)}{\kappa\,f}\right]^{3/f},\;\;\;\mbox{and}\;\;
\beta_1=\frac{2(4f+3)}{f}.
$$
%From Eq.(\ref{pot}) we obtain that the scalar potential $V(\phi)$
%is given by
%\begin{equation}
%V(\phi)=C\phi^{-\beta_2}+D\phi^{-\beta_3},\label{pot1}
%\end{equation}
%where
%$$
%C=\frac{3\,f^2\,A^2}{\kappa}\,\left[
%\frac{f\,\kappa}{8\,A\,(1-f)}\right]^{2(f-1)/f},\,\;\;\beta_2=\frac{4(1-f)}{f},
%$$
%$$
%D=\frac{f\,A\,(f-1)}{\kappa}\,\left[
%\frac{f\,\kappa}{8\,A\,(1-f)}\right]^{(f-2)/f},\;\;\;\mbox{and}
%\,\,\;\beta_3=\frac{2\,(2-f)}{f}.
%$$
Using the slow-roll approximation, $\dot{\phi}^2/2<V(\phi)$, and
$V(\phi)>\rho_\gamma$, the scalar potential given by
Eq.(\ref{pot}) reduces to
\begin{equation}
V(\phi)\simeq \frac{3\,H^2}{\kappa}=C\phi^{-\beta_2},\label{pot11}
\end{equation}
where
$$
C=\frac{3\,f^2\,A^2}{\kappa}\,\left[
\frac{f\,\kappa}{8\,A\,(1-f)}\right]^{2(f-1)/f},\,\;\;\beta_2=\frac{4(1-f)}{f}.
$$
Note that this kind of potential does not present a minimum.  Note
also that the scalar field $\phi$, the Hubble parameter $H$, and
the potential $V(\phi)$ become independent of the parameters
$C_\phi$ and $C_\gamma$.

Introducing the dimensionless slow-roll parameter $\varepsilon$,
we write
\begin{equation}
\varepsilon=-\frac{\dot{H}}{H^2}=\frac{8\,(1-f)^2}{\kappa\,f^2}\;\frac{1}{\phi^2},\label{ep}
\end{equation}
and the second slow-roll parameter  $\eta$
\begin{equation}
\eta=-\frac{\ddot{H}}{H
\dot{H}}=\frac{8\,(1-f)\,(2-f)}{\kappa\,f^2}\;\frac{1}{\phi^2}\,.\label{eta}
\end{equation}

So, the condition for inflation to occur  $\ddot{a}>0$ (or
equivalently $\varepsilon<$1)   is only satisfied when
$\phi^2>\frac{8\,(1-f)^2}{\kappa\,f^2}$.

%Note that the ratio between  $\varepsilon$ and $\eta$ becomes
%$\varepsilon/\eta=\frac{(1-f)}{(2-f)}$ and  thus $\eta$ is always
%larger than $\varepsilon$, since $0<f<1$. Note, also, that $\eta$
%reaches unity before $\varepsilon$ does. In this way, we may
%establish that the end of  inflation is governed by the condition
%$\eta=1$ in place of $\varepsilon=1$. From this condition we get
%for scalar field, at the end of inflation the value
%\begin{equation}
%\phi_{end}=\frac{2}{f}\;\left[\frac{2\,(1-f)\,(2-f)}{\kappa}\right]^{1/2}.
%\label{al}
%\end{equation}

The number of e-folds between two different values of cosmological
times $t_1$ and $t_2$ (or equivalently between two values $\phi_1$
and $\phi_2$ of the scalar field)   is given by
\begin{equation}
N=\int_{t_1}^{t_{2}}\,H\,dt=A\,(t_{2}^f-t_1^f)=\,\frac{f\,\kappa}{8\,(1-f)}
(\phi_{2}^{2}-\phi_1^{2}).\label{N1}
\end{equation}
Here we have used Eq.(\ref{wr1}).

If we assume that inflation begins at the earliest possible stage,
that it, at $\varepsilon=1$ (or equivalently $\ddot{a}=0$ ), the
scalar field becomes
\begin{equation}
\phi_{1}=\frac{2\,\sqrt{2}\,(1-f)}{f\,\sqrt{\kappa}}\;. \label{al}
\end{equation}

%  In the following, the subscripts  $*$ and $end$ are
%used to denote  the epoch when the cosmological scales exit the
%horizon and the end of  inflation, respectively.

As argued in Refs.\cite{warm,Liddle}, the density perturbation
could be written as
${\cal{P}_{\cal{R}}}^{1/2}=\frac{H}{\dot{\phi}}\,\delta\phi$. In
particular in the warm inflation regime, a thermalize radiation
component is present, therefore, inflation fluctuations are
dominantly thermal rather than quantum.  In the weak dissipation
limit, we have $\delta\phi^2\simeq H\,T$ \cite{62526,B1}. From
Eqs.(\ref{inf3}) and (\ref{rh-1}),  ${\cal{P}_{\cal{R}}}$ becomes
\begin{equation}
{\cal{P}_{\cal{R}}}\simeq\left[\frac{\kappa^3\,\Gamma}{2^5\,C_\gamma}\right]^{1/4}\;
\left[\frac{H^{11/3}}{-\dot{H}}\right]^{3/4}=
\beta_4\;\phi^{2(f-2)/f}, \label{pd}
\end{equation}
where
$$
\beta_4=\left[\frac{B\,\kappa^{3}}{2^5\,C_\gamma}\right]^{1/4}\;B_1
,\;\;\;\mbox{and}\;\;\;B_1=\frac{f^2\,A^2}{(1-f)^{3/4}}\;\left[\frac{\kappa\,f}{8\,
A\,(1-f)}\right]^{(8f-5)/(4f)}.
$$
The scalar spectral index $n_s$ is given by $ n_s -1 =\frac{d
\ln\,{\cal{P}_R}}{d \ln k}$,  where the interval in wave number is
related to the number of e-folds by the relation $d \ln k(\phi)=d
N(\phi)=(H/\dot{\phi})\,d\phi$. From Eqs. (\ref{wr1}) and
(\ref{pd}), we get,
\begin{equation}
n_s=1-\frac{8(1-f)\,(2-f)}{\kappa\,f^2}\;\frac{1}{\phi^2}.\label{nss1}
\end{equation}
Since $1>f>0$, we clearly see that the scalar index in the weak
dissipative regime becomes $n_s<1$. The scalar spectral index can
be re-expressed in terms of the number of e-folding, $N$. By using
Eqs.(\ref{N1}) and (\ref{al}) we have
\begin{equation}
n_s=1-\frac{(2-f)}{[1+f\,(N-1)]},
\end{equation}
and the value of $f$ in terms of the $n_s$ and $N$ becomes
$$
f=\frac{(1+n_s)}{N(1-n_s)+n_s}.
$$
In particular, for $n_s=0.96$ and $N=60$ we obtain that $f\simeq
0.58$.

From Eqs.(\ref{N1}), (\ref{al}), (\ref{pd}) and  (\ref{nss1}), we
can write the parameter $A$ in terms of the particle physics
parameters $C_\gamma$ and $C_\phi$, and ${\cal{P}_R}$, $N$ and
$n_s$ (since, $f$  is function of the $N$ and $n_s$, as could be
seen from the latter equation), in the form

\begin{equation}
A=\left(\frac{C_\gamma}{C_\phi}\right)^{f/2}\,\frac{{\cal{P}_R}^{f/2}\;
B_2}{[1+(N-1)\,f]^{(f-2)/2}},\label{A}
\end{equation}
where $B_2$ is given by
$$
B_2=(108)^f\,f^{-(f+1)}\,\left[\frac{8\,(1-f)}{\kappa}\right]^{f/2}.
$$

\begin{figure}[th]
\includegraphics[width=6.0in,angle=0,clip=true]{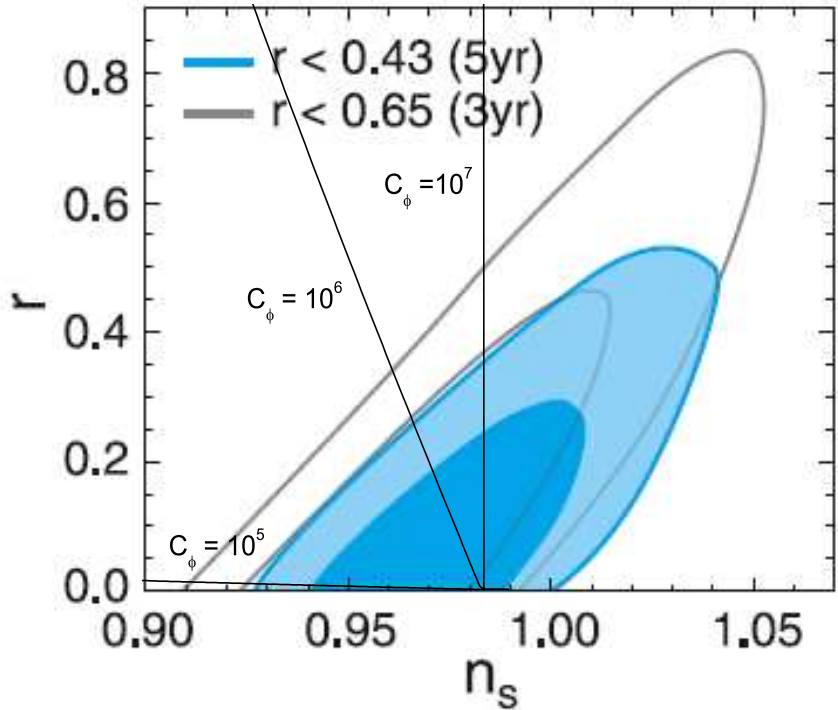}
\caption{Evolution of the tensor-scalar ratio $r$ versus the
scalar spectrum index $n_s$ in the weak dissipative regime, for
three different values of the parameter $C_\phi$ . Here we used
$f=1/2$, $\kappa=1$, $C_\gamma=70$, and
${\cal{P}_{\cal{R}}}=2.4\times 10^{-9}$.
 \label{rons}}
\end{figure}

As it was mentioned in Ref.\cite{Bha} the generation of tensor
perturbations during inflation would  produce  gravitational wave.
The corresponding spectrum  becomes
${\cal{P}}_g=8\kappa(H/2\pi)^2$. For $R<1$ and from Eq.(\ref{pd})
we may write the tensor-scalar ratio as
\begin{equation}
r(k)=\left(\frac{{\cal{P}}_g}{P_{\cal R}}\right)
\simeq\left[\frac{\beta_5\;\phi^2}{\beta_4}\right],
\label{Rk}\end{equation} where
$$
\beta_5=\frac{8\kappa\,A^2\,f^2}{4\,\pi^2}\;\left[\frac{f\,\kappa}{8\,A\,(1-f)}\right]^{2(f-1)/f},
$$
and $r$ in terms of the scalar spectral index, becomes
\begin{equation}
r\simeq\left[\frac{\beta_5}{\beta_4}\right]\,\left[\frac{8\,(1-f)\,(2-f)}{\kappa\,f^2\,(1-n_s)}\right].
\label{Rk1}
\end{equation}

Analogously, we can write the tensor-scalar ratio as function of
the number of e-folding

\begin{equation}
r
\simeq\left[\frac{8\,(1-f)\,\beta_5}{\kappa\,f^2\,\beta_4}\right]\;[1+f(N-1)].
\label{Rk11}\end{equation}

In Fig.(\ref{rons}) we show the dependence of the tensor-scalar
ratio on the spectral index for the special case in which we fixe
$f=1/2$, and we have used three  different values of the parameter
$C_\phi$ . From left to right $C_\phi$=$10^5$, $10^6$ and $10^7$.

%From Ref.\cite{astro}, two-dimensional marginalized constraints
%(68$\%$ and 95$\%$ confidence levels) on the inflationary
%parameters $r$  and $n_s$, defined at $k_0$ = 0.002 Mpc$^{-1}$.

The five-year WMAP data places stronger limits on $r$ (shown in
blue) than three-year data (grey)\cite{Spergel}. In order to write
down values that relate $n_s$ and $r$, we used Eq.(\ref{Rk1}) .
Also we have used the values $C_\gamma=70$ and $\kappa=1$. Note
that for the value of the parameter $C_\phi$, (restricted from
below, in which $C_\phi>10^5$),   is well supported by the data.
From Eqs.(\ref{A}) and (\ref{Rk11}), we observed that for the
special case in which $C_\phi=10^6$  and $f = \frac{1}{2}$, the
curve $r = r(n_s)$ for WMAP 5-year enters the 95$\%$ confidence
region for $r\simeq 0.26$, which corresponds to the number of
e-folds, $N \simeq 146$. For $r \simeq 0.20$ corresponds to $N
\simeq 140$, in this way the model is viable for large values of
the number of e-folds.

\section{ The  strong dissipative regime.\label{section3}}

We consider now the case in which $\Gamma$ is large enough for the
system to remain in strong dissipation until the end of inflation,
i.e. $R>1$. From Eqs.(\ref{inf3}) and (\ref{at}), we can obtained
a relation between the scalar field and cosmological times given
by
\begin{equation}
\ln\left[\frac{\phi(t)}{\phi_0}\right]=\,\alpha_1\;t^{(5f+2)/8},\label{wr12}
\end{equation}
where $\phi(t=0)=\phi_0$ and $\alpha_1$ is defined by
$$
\alpha_1=\frac{8}{(5f+2)}\,(A\,f)^{5/8}\,(1-f)^{1/8}\,\sqrt{\frac{6}{\kappa\,\alpha}},\;\;\;\mbox{and
}
\;\;\;\alpha=C_\phi\,\left[\frac{2}{3\,\kappa\,C_r}\right]^{3/4}.
$$
The Hubble parameter as a function of the inflaton field, $\phi$,
result as
\begin{equation}
H(\phi)=A\,f\,\left[\frac{1}{\alpha_1}\;\ln(\phi/\phi_0)\right]^{-8(1-f)/(5f+2)}.\label{HH2}
\end{equation}
Without loss of generality we can taken $\phi_0=1$.

From Eq.(\ref{G1}) the dissipation coefficient, $\Gamma$, can be
expressed as a function of the scalar field, $\phi$, in the case
of the strong dissipation regime, as follows
\begin{equation}
\Gamma(\phi)=\frac{\alpha_2}{\phi^2}\,\left[\ln(\phi)\right]^{-\alpha_3},\label{gg2}
\end{equation}
where
$$
\alpha_2=\alpha\,[A\,f\,(1-f)]^{3/4}\;\alpha_1^{\alpha_3},\;\;\;\mbox{and}\;\;\;\,\alpha_3=\frac{6(2-f)}{(5f+2)}.
$$
Analogously, to the case of the weak dissipative regime, we can
used the slow-roll approximation i.e. $\dot{\phi}^2/2<V(\phi)$,
together with $V(\phi)>\rho_\gamma$. From Eq.(\ref{pot}) the
scalar potential becomes
\begin{equation}
V(\phi)\simeq
\left(\frac{3\,f^2\,A^2}{\kappa}\right)\;\left[\frac{1}{\alpha_1}\;\ln(\phi)\right]^{-16(1-f)/(5f+2)},\label{pot11}
\end{equation}
and as in the previous case,  this kind of potential does not
present a minimum.

Introducing the dimensionless slow-roll parameter $\varepsilon$,
we have
\begin{equation}
\varepsilon=-\frac{\dot{H}}{H^2}=\left(\frac{1-f}{f\,A}\right)\;\left[\frac{\alpha_1}{\ln(\phi)}\right]^{8f/(5f+2)},\label{ep1}
\end{equation}
and the second slow-roll parameter  $\eta$ is given by
\begin{equation}
\eta=-\frac{\ddot{H}}{H
\dot{H}}=\left(\frac{2-f}{f\,A}\right)\;\left[\frac{\alpha_1}{\ln(\phi)}\right]^{8f/(5f+2)},\label{eta2}
\end{equation}

%Again  $\eta$ reaches unity before $\varepsilon$ does. In this
%way, we may establish that the end of  inflation is governed by
%the condition $\eta=1$.
Imposing  the condition $\varepsilon=1$  at the beginning of
inflation the scalar field, $\phi$, takes at this time the value
\begin{equation}
\phi_{1}=\exp\left(\alpha_1\;\left[\frac{1-f}{f\,A}\right]^{\frac{(5f+2)}{8f}}\right).
\label{al22}
\end{equation}

The number of e-folds  becomes given by
\begin{equation}
N=\int_{t_1}^{t_{2}}\,H\,dt=A\,\alpha_1^{-8f/(5f+2)}\,
[\ln(\phi_{2})^{8f/(5f+2)}-\ln(\phi_{1})^{8f/(5f+2)}],
\label{N22}
\end{equation}
where Eq.(\ref{wr12}) was used.

In the case of high dissipation,  i.e. $R=\Gamma/3H\gg 1$ and
following  Taylor and Berera\cite{Bere2}, we can write
$\delta\phi^2\simeq\,\frac{k_F\,T\,}{2\,\pi^2}$, where  the
wave-number $k_F$ is defined by $k_F=\sqrt{\Gamma H}=H\,\sqrt{3
R}> H$, and corresponds to the freeze-out scale at which
dissipation damps out to the thermally excited fluctuations. The
freeze-out time is defined by the condition $k=a\,k_F$.
%wave-number
%$k_F$ is defined at the point where the inequality
%$V_{,\,\phi\,\phi}< \Gamma H$, is satisfied \cite{Bere2}.
From Eqs.(\ref{wr12}) and (\ref{gg2}) we obtained that
\begin{equation}
{\cal{P}_{\cal{R}}}\simeq\frac{1}{2\,\pi^2}\,
\left[\frac{\Gamma^{3}\,H^9}{4\,C_\gamma\,\dot{\phi}^6}\right]^{1/4}\simeq\frac{1}{4\,\pi^2}\,
\left[\frac{\kappa^3\,\Gamma^{6}\,H^6}{54\,C_\gamma\,(-\dot{H})^3}\right]^{1/4}\simeq\alpha_4\,\frac{\ln(\phi)^{\alpha_5}}{\phi^3}\,,
\label{pd21}
\end{equation}
where
$$
\alpha_4=\frac{1}{4\pi^2}\,\left[\frac{\kappa^3\,\alpha_2^{6}\,A^6\,f^6}
{54\,C_\gamma\,[A\,f\,(1-f)]^3}\right]^{1/4}\,\alpha_1^{-6\,f/(5f+2)},\,\:\;\mbox{and}\;\;
\alpha_5=\frac{(15f-18)}{(5f+2)}.
$$
From Eqs.(\ref{wr12}) and (\ref{gg2}) the scalar spectral index
$n_s=d\,{\cal{P}_{\cal{R}}}/d\ln k$, is given by
\begin{equation}
n_s\simeq
1-\left[\frac{3\,\ln(\phi)-\alpha_5}{\alpha_6\,\ln(\phi)^{\alpha_7}}\right],\label{nss}
\end{equation}
where
$$
\alpha_6=\frac{(A\,f)^{3/8}}{(1-f)^{1/8}}\,\sqrt{\frac{\kappa\,\alpha}{6}}\,\,\alpha_1^{(2-3\,f)/(5\,f+2)},
\;\;\mbox{and}\,\;\;\;\alpha_7=\frac{8\,f}{(5\,f+2)}.
$$
The scalar spectra index $n_s$ also can be write in terms of the
number of e-folds $N$. Thus, using Eqs. (\ref{al22}) and
(\ref{N22}), we get
\begin{equation}
n_s\simeq
1-f\,A\,\left[\frac{3\,\alpha_1\,[1+f\,(N-1)]^{-\alpha_7}\,(f\,A)^{\alpha_7}-\alpha_5
}{\alpha_6\,[1+f(N-1)]\,\alpha_1^{\alpha_7}} \right].\label{ns2}
\end{equation}

For the strong dissipative regime  we may write the tensor-scalar
ratio as
\begin{equation}
r(k)=\left(\frac{{\cal{P}}_g}{P_{\cal R}}\right)
\simeq\,\left[\frac{2\,\kappa\,(f\,A)^2}{\pi^2\,\alpha_4}\,\alpha_1^{16(1-f)/(5f+2)}\right]\;\phi^3\;[\ln(\phi)]^{(f+2)/(5f+2)}.
\label{Rk2}
\end{equation}
Here, we have used expressions (\ref{HH2}) and (\ref{pd21}).

\begin{figure}[th]
\includegraphics[width=6.0in,angle=0,clip=true]{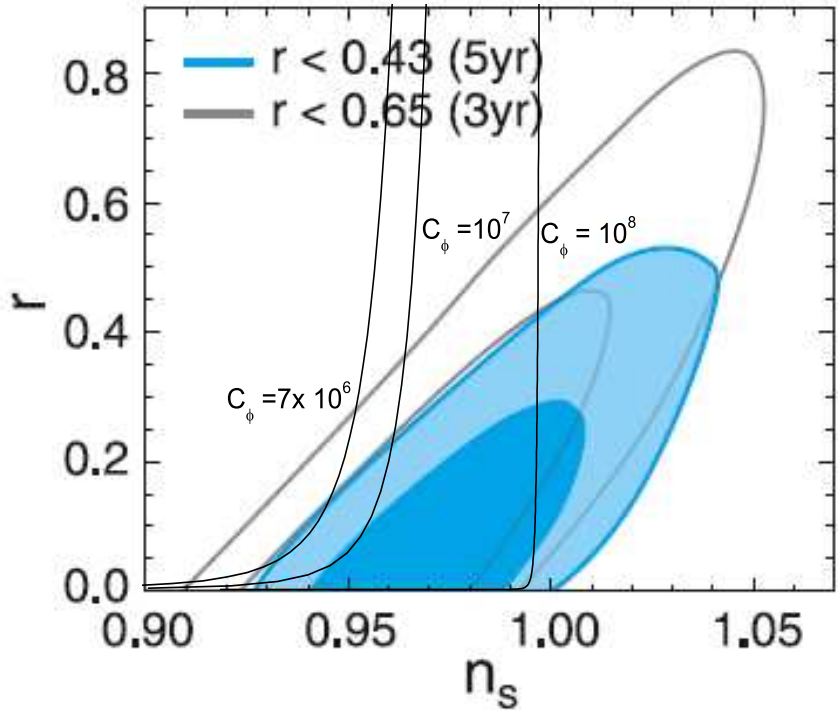}
\caption{Evolution of the tensor-scalar ratio $r$ versus the
scalar spectrum index $n_s$ in the strong dissipative regime, for
three different values of the parameter $C_\phi$. Here, we have
used $f=1/2$, $\kappa=1$, $C_\gamma=70$  and
${\cal{P}_{\cal{R}}}=2.4\times 10^{-9}$.
 \label{fig2}}
\end{figure}

Fig.(\ref{fig2})  shows (for the strong dissipative regime) the
dependence of the tensor-scalar ratio on the spectral index. Here,
we have used  different values for the parameter $C_\phi$.
% From
%Ref.\cite{astro}, two-dimensional marginalized
% constraints (68$\%$ and 95$\%$ confidence levels) on inflationary parameters
%$r$, the tensor-scalar ratio, and $n_s$, the spectral index of
%fluctuations, defined at $k_0$ = 0.002 Mpc$^{-1}$.
The  WMAP five-year data places stronger limits on $r$ (shown in
blue) than three-year data (grey)\cite{Spergel}. In order to write
down values that relate $n_s$ and $r$, we used Eqs.(\ref{pd21}),
(\ref{nss}) and (\ref{Rk2}). Also we have used the WMAP value
$P_{\cal R}(k_*)\simeq 2.4\times 10^{-9}$, the value $f=0.5$ and
$C_\gamma=70$. Note that for  the values of the parameter $C_\phi$
greater than $5\times 10^{6}$, our model is well supported by the
data. From Eqs.(\ref{al22}), (\ref{N22}), (\ref{pd21}) and
(\ref{Rk2}), we observed numerically that for the special case in
which we fixes $C_\phi=10^{7}$ and $f = \frac{1}{2}$, the curve $r
= r(n_s)$ for WMAP 5-years enters the 95$\%$ confidence region for
$r\simeq 0.245$, which corresponds to the number of e-folds, $N
\simeq 359$.  For $r \simeq 0.11$ corresponds to $N \simeq 289$,
in this way the model is viable for large values of the number of
e-folds.

\section{Conclusions \label{conclu}}

In this paper we have studied the warm-intermediate inflationary
model in the weak and  strong dissipative regimes.  In the
slow-roll approximation we have found a general relation between
the scalar potential and its derivative. We have also obtained
explicit expressions for the corresponding, power spectrum of the
curvature perturbations $P_{\cal R}$, tensor-scalar ratio $r$,
scalar spectrum index $n_s$ and the number of e-folds $N$.

In order to bring some explicit results we have taken the
constraint $r-n_s$ plane to first-order in the slow roll
approximation. When $\Gamma<3H$ warm inflation occurs in the
so-called weak dissipative regimen. In this case, the dissipation
coefficient $\Gamma\propto\phi^{-\beta_1}$ for intermediate
inflation. We also noted that the parameter $C_\phi$, which is
bounded from bellow,  $C_\phi>10^5$, the model is well supported
by the data as could be seen from Fig.(\ref{rons}). Here, we have
used the WMAP five year data, where $P_{\cal R}(k_*)\simeq
2.4\times 10^{-9}$, and we have taken the value $f=1/2$. On the
other hand, when $\Gamma>3H$ warm inflation occurs in the
so-called strong dissipative regime. In this regime, the
dissipation coefficient $\Gamma$ present a dependence proportional
to $(\log(\phi))^{-\alpha_3}/\phi^2$ in intermediate inflation. In
particular, Fig.(\ref{fig2}) shows that for the values of the
parameter $C_\phi= 7\times 10^6$, $10^7$ or $10^8$, the model is
well supported  by the WMAP data, when the value $f=1/2$ is taken.

In this paper, we have not addressed the non-Gaussian effects
during warm inflation (see e.g., Refs.\cite{27,fNL}). A possible
calculation from the non-linearity parameter $f_{NL}$, would give
new constrains on the parameters of the model. We hope to return
to this point in the near future.

\begin{acknowledgments}
 This work was
supported by COMISION NACIONAL DE CIENCIAS Y TECNOLOGIA through
FONDECYT grants N$^0$ 1070306 (SdC) and N$^0$ 1090613 (RH and
SdC).
\end{acknowledgments}

%\\\\\\\\\\\\\\\\\\\\\\\\\\\\\\\\\\\\\\\\\\\\\\\\\\\\\\\\\\\\\\\\\\\\\\\

\end{document}